\documentclass[12pt]{article}
\usepackage[dvips]{graphicx}
\begin{document}

\author{D.~Lacroix and Ph. Chomaz \\
%EndAName
G.A.N.I.L, B.P. 5027, F-14076 Caen Cedex 5, France. }
\title{Beating of monopole modes in nuclear dynamics.}
\maketitle

\begin{abstract}
Time-dependent Hartree-Fock simulations of the evolution of excited gold
fragments have been performed. The observed dynamics appears more complex
than the collective expansion picture. The minimum density is often not
reached during the first density oscillation because of the beating of
several collective compression modes.
\end{abstract}

\section{Introduction}

Studying the dynamics of a highly excited nucleus, such as those created in
Heavy-ion collisions, is a priori a hard task since we are dealing with a
many-body quantal problem involving a large number of degrees of freedom. A
widely used approximation is to reduce the problem to the dynamics of few
collective variables, the remaining degrees of freedom being treated in a
statistical manner. In particular, excited nuclear systems are often assumed
to simply be hot nuclei expanding or contracting isentropically in a unique
collective compression mode, the so-called breathing mode ( see for example
the expanding fireball scenario or the fragmentation models \cite
{Ber1,Cug1,Fri1}).

This picture has been recently applied to explain the GSI data concerning
the correlation between the apparent temperature and the excitation energy
of fragmentation events known as the Aladin ''caloric-curve'' (see ref.\cite
{Papp1} and refs. therein). In the approach of ref.\cite{Papp1} the
fragmentation of various nuclear systems formed by the projectile remains
after a collision, is assumed to occur at the respective turning points of
the induced giant breathing mode. The location of these turning points in
temperature and density is proposed to be the explanation of the observed
''caloric-curve''\cite{ALADIN}. More generally, most of the
multifragmentation data are understood assuming that the system expands to
reach a low density value either in statistical models\cite{stat}
or in dynamical simulations\cite{spinodal} using
stochastic approaches\cite{Ayi88,Ran90}. The general belief is that
the system breaks during the first expansion associated with the giant
breathing mode of the hot nucleus. More generally, this assumption is often
made in simple semi-classical simulations or in multifragmentation models 
\cite{Ber1,Cug1,Fri1}.

However, recent results about photon production during Heavy-Ion collisions
seems not to be compatible with this simple scenario of a fast
multi-fragmentation during the first expansion of the system and the data
have been understood assuming at least one recompression before break-up\cite
{Mar95,Sch1}.

On the other hand, from the study of monopole vibrations, it is well-known
that nuclei can exhibit a large variety of compression modes\cite
{beating,monopole}. Moreover, microscopic description (using TDHF or is
linear version) of monopole vibrations predicts that these modes cannot be
simply described by the breathing mode picture. Since these giant resonances
are the small amplitude limit of the collective expansion assumed to occur
in heavy ion collisions, one may worry if the picture of a unique large
amplitude collective breathing mode leading the hot nuclei to low density
regions is not too simple. Therefore, in this article, we propose to study
the large amplitude motion of an excited nucleus within the framework of the
Time-Dependent Hartree-Fock (TDHF) theory in order to critically discuss the
validity of the single breathing mode picture. It should be noticed that,
heavy ion collisions are rather well described by simulations using the
semi-classical version of TDHF: the Vlasov equation, possibly extended to
take into account the damping coming from nucleon-nucleon collisions (i.e.
the so-called BUU approach\cite{Bertch}). In these simulations the expansion
phase appears to be dominated by the mean-field dynamics confirming that
TDHF can be a good tool to investigate this large amplitude motion. 
Furthermore, as far as we know, TDHF is the only well-tested quantum model
tractable for Heavy-Ion collisions. Of course, a quantum model that include
collision effects (extended TDHF) would be preferable. However, we believe
that the introduction of this effect will not change considerably the
discussion since, during the first expansion, the gain and lost term of
two-body collisions are almost equal and so cancel.

In this article, we will analyzed the TDHF dynamics and we will point out that
some aspects, important for the dynamics, cannot be accounted for in the
simple breathing mode picture. This may change our understanding of the
expansion dynamics of nuclei created during heavy ion collisions. In
particular, we observe that, in general, the minimum central density
is not reached during the first density oscillation. This may allow the
system to undergo several monopolar oscillations before breaking in pieces
as the photon data are suggesting.

\section{Mean-Field dynamics}

In mean-field evolution all the information is contained in the
single-particle density operator $\hat{\rho}$. Let us introduce the
single-particle basis which diagonalizes this operator: $\hat{\rho}=\sum {%
|\Phi _\beta >n_\beta <\Phi _\beta |}$, where $\beta $ contains all quantum
numbers. Following the work of ref.\cite{Vau1}, we consider the mean-field
evolution of a spherical symmetric\footnote{ Either experimental
evidence, for example in central events selected with INDRA, or theoretical
simulation of the BUU type confirms the fact that a spherical equilibrated
source can be a reasonable ansatz.}, spin-isospin saturated nucleus. The
mean-field is parameterized as

\begin{equation}
U(r)=\frac{3}{4}t_{0}\rho (r)+\frac{(\sigma +2)}{16}{\rho (r)}^{\sigma
+1}+V_{0}\int {\frac{\exp {\frac{|r-r^{\prime }|}{a}}}{\frac{|r-r^{\prime }|%
}{a}}\rho (r^{\prime })}d{\ r^{\prime }}+V_{C}  \label{eq:MF}
\end{equation}
The two first terms are the same as in ref.\cite{Vau1} and a general
discussion could be found in \cite{Vau2}. The Yukawa potential is introduced
in order to mimic the gradient part of the potential as in ref.\cite
{Koo1,Bon1}. Finally, $V_{C}$ is the direct Coulomb potential which is ,in
our calculation, approximately treated by given a charge $\frac{Z}{A}$ to
each nucleon. Parameters are chosen in order to have the same property as
the $SkM^{*}$\cite{Gue1} force for the nuclear matter. We have taken $%
t_{0}=-2191.73$ MeV fm$^{3}$, $t_{3}=18818.8$ MeV fm$^{7/2}$, $\sigma =\frac{
1}{6}$, $a=0.45979$ fm, $V_{0}=-461.07$ MeV fm$^{3}$. For the nuclear
matter, this leads to $E/A=-15.8$ MeV at a saturation density $\rho _{\infty
}=0.16$ fm$^{-3}$ and to an incompressibility modulus of $K_{\infty }=200$
MeV which corresponds to a soft Equation Of State. It should be noticed that
the presented results do not qualitatively depend upon the compressibility
of the nuclear force.

Since in TDHF, the initial symmetry of the system is conserved, the
dynamical evolution reduces to a set of equations for the radial part $%
\varphi_{nl}(r,t)$ of the single-particle wave functions $\Phi _{\beta }(%
\vec{r},t)$

\begin{equation}
i\hbar \frac{\partial }{\partial {t}}\varphi _{nl}(r)=\left\{ \frac{-\hbar
^{2}}{2m}\frac{\partial }{\partial {r^{2}}}+\frac{\hbar ^{2}l(l+1)}{2m r^{2}}
+U(\rho (r,t))\right\} \varphi _{nl}(r)  \label{eq:dynamic}
\end{equation}
where $n$ and $l$ are the principal and orbital numbers and the density
takes the particular form

\begin{equation}
\rho (r,t)=4\sum_{n,l}(2l+1)n_{nl}\frac{\left| \varphi _{nl}(r,t)\right| ^{2}%
}{4\pi r^{2}},  \label{eq:Rho}
\end{equation}

Numerically, we consider $40$ orbitals, $\varphi _{nl}$. The Schr\"{o}dinger
equation is solved in coordinate representation. The lattice of size $300$
fm is discretized in steps of size $\Delta r=0.2$ fm. An imaginary time
method is used to generate the ground state of the static problem \cite
{Vau1,Vau2}( a small external field $\lambda r^2$ with $\lambda =0.25$ MeV fm%
$^{-2}$ is added during this initialization in order to define the particle
states). Occupation numbers are calculated afterwards according to the
Fermi-Dirac distribution at the considered temperature. We perform the TDHF
evolution up to $1500$ fm/c with a time step of $0.75$ fm/c.

In order to study nuclei comparable to those expected to be produced in
reactions, we take the same initial conditions as in ref.\cite{Papp1} for
the fragments of a Au-projectile. Let us first focus on one typical initial
condition : an excited fragment of mass $A=191.$ The excitation energy is
taken from the abrasion-ablation model\cite{Gaim1}. Here we consider that
the excitation energy is related to the number ($\Delta A)$ of particles
lost during the abrasion-ablation stage of the collision by the relation $%
E^{*}=13.3\Delta A$ MeV. Furthermore, following ref.\cite{Papp1} we assume
that the projectile fragments are initially slightly dilated. We implement
this dilution by rescaling the wave-function according to $r\rightarrow
r\;0.8^{-1/3}$. In this condition the temperature is adjusted in order to
obtain the correct excitation energy. For the considered fragment which is
closed from the projectile, the excitation energy is small and partly stored
in the initial dilution so that the temperature is only $T=0.55MeV$. As we
will discuss in the following, the conclusion we will draw are rather
general and the complex dynamics is observed for a wide range of
initial value of the mass, the density and the temperature.

\section{Results}

\subsection{Density profiles}

\begin{figure}[tbph]
\begin{center}
\includegraphics*[height=15cm,width=8cm]{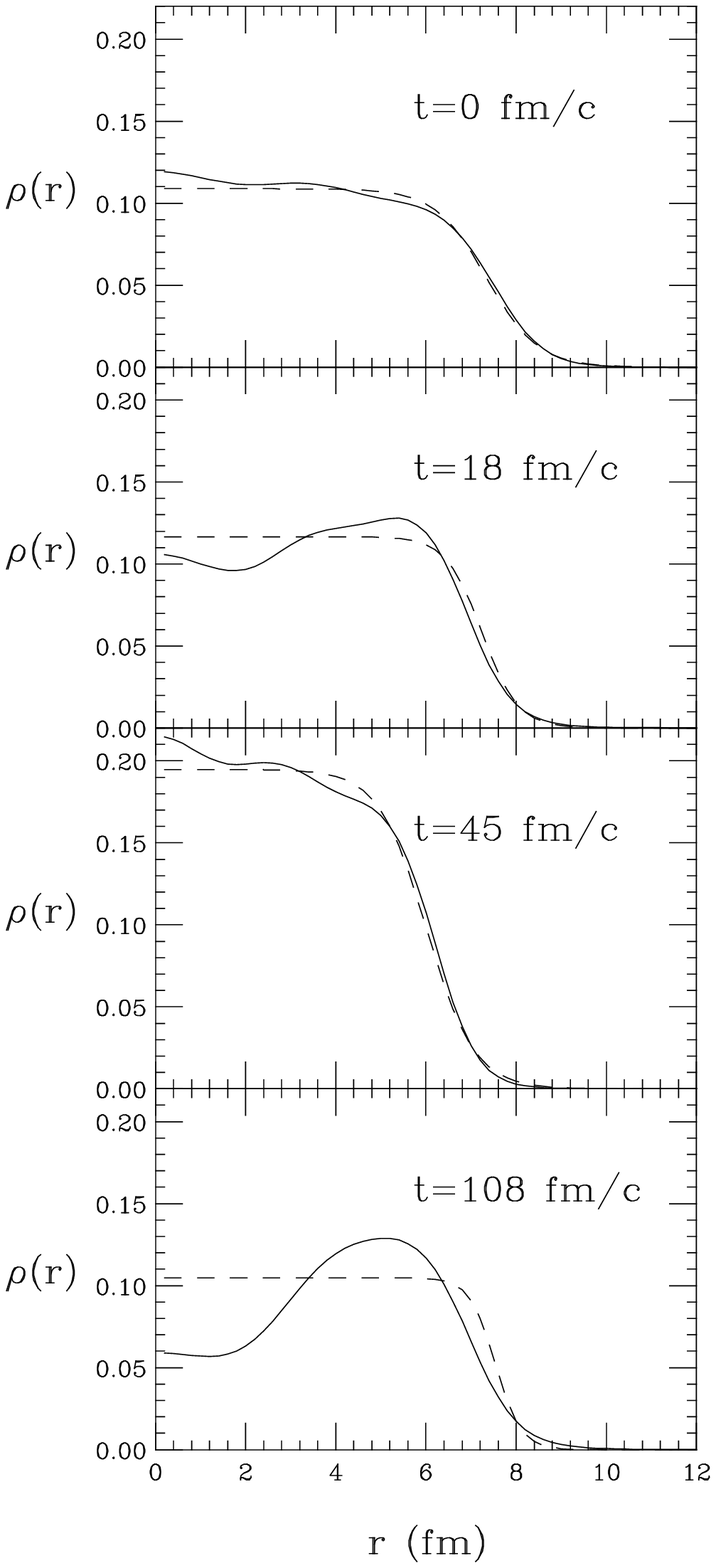}
\end{center}
\caption{Density profile (solid curve) for different time. Dashed curve
represent fits with a Fermi shape parametrisation of the density. This
fit is not able to reproduce the complicated density profile (see $t=108 fm/c
$) and demonstrate the difference between the TDHF evolution and the
scaling picture.}
\label{fig:1}
\end{figure}

In fig.\ref{fig:1} we display the density (solid line) at various times as
predicted by the mean-field approximation\footnote{%
It should be noticed that we have checked that putting the ingredients
considered in reference [13] we recover their results. The difference is
that we are making a more detailed analysis for a heavier nucleus.}. We
observe that the density profile display a lot of distortions. To emphasize
this point we have fitted the density profile at each time by a Fermi shape

\begin{equation}
\rho _{_{F}}(r,t)=\frac{\rho (t)}{1+\exp \left( \frac{r-R(t)}{a(t)}\right) }
\label{eq:wood}
\end{equation}
Typical results of the fit are displayed in fig.\ref{fig:1}(dashed curves). 
At initial time, the density profile is well fitted by a Fermi shape.
In a scaling picture of the evolution, one expects, at all time, that this
fit should be also good. The strong difference between the fitted Fermi
shape and the real density profile at time $t=108$ fm/c demonstrates how far
we are from a simple scaling picture of the expansion.

From the three parameters $\rho (t)$, $R(t)$ and the diffuseness coefficient 
$a(t)$, we can define three scaling parameters as

\begin{equation}
\alpha _{R}=\frac{R(t)}{R(t_{0})}\hspace{1cm}\hspace{1cm}\alpha _{a}=\frac{
a(t)}{a(t_{0})}\hspace{1cm}\hspace{1cm}\alpha _{\rho }=\left( \frac{\rho
(t_{0})}{\rho (t)}\right) ^{\frac{1}{3}}  \label{eq:scaling}
\end{equation}
where $t_{0}$ is the initial time. If the collective dynamics is dominated
by a unique breathing mode (a global scaling of the density profile) the
three $\alpha $ should be equal and should oscillate with a unique frequency
close to the one of the monopolar vibration of the considered nucleus.%
\footnote{%
A possible shift in frequency can be due to the anharmonicity of the large
amplitude motion and to the effect of the temperature.}

\begin{figure}[tbph]
\begin{center}
\includegraphics*[height=10cm,width=10cm]{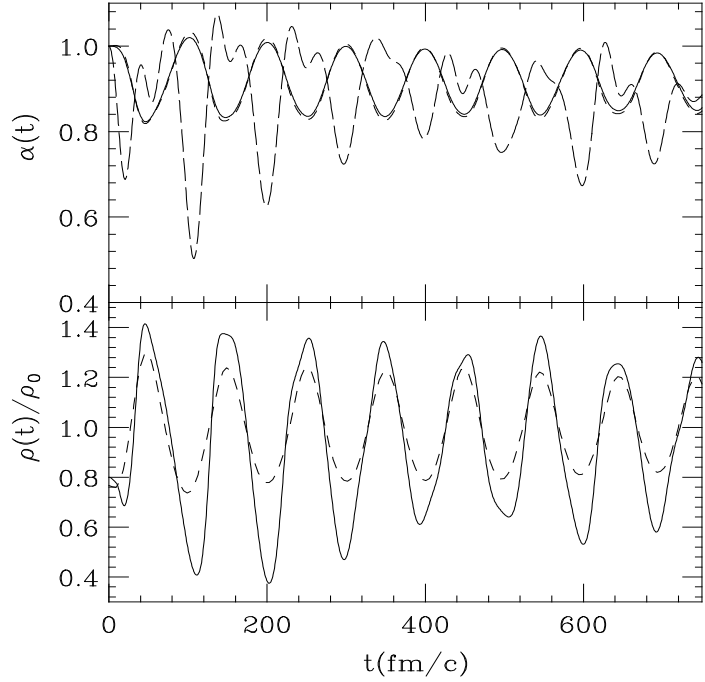}
\end{center}
\caption{Top: Evolution of the three scaling factor calculated with the
three parameters of the Fermi shape density. Dashed line: $\alpha _{a}$ ,
long-dashed line: $\alpha _{R}$ and solid line: $\alpha _{\rho }$. Bot:
Evolution of the central densitie $\rho_{2fm}$ (solid curve) and $\rho_{5fm}$
(dashed curve) as a function of time.}
\label{fig:2}
\end{figure}

In fig.\ref{fig:2} (Top), we show the evolution of the $\alpha $'s as a
function of time. From this figure, it is clear that only $\alpha _{R}$ and $%
\alpha _{\rho }$ display regular oscillation with a given frequency.
Performing a Fourier transform of this quantities, we found a marked peak at 
$\hbar \omega =12.5$ MeV (which is around the expected breathing mode
frequency of the considered nucleus at zero temperature).

On contrary, $\alpha _a$ contains clearly a complex superposition of many
modes and is never comparable to the two other scaling factors. The Fourier
transform of $\alpha _a$ presents not only the frequency $\hbar \omega =12.5$
MeV but also two other broad peaks around $17$ MeV and $26$ MeV, and a small
peak around $40$ MeV. The Fermi shape fit being unrealistic for the
density profile at time $t=108fm/c$, it is difficult to really conclude on
the link between the presence of different collective modes and the
appearance of hollow shapes. In the following, we will characterize more
precisely this link.

From fig. \ref{fig:1} it is clear that not only the surface of the nucleus
is following a complex dynamics but also the interior which exhibits a
tendency to produce hollow structures. To quantify this feature, we have
computed two different averaged densities as a function of time: the solid
line in figure \ref{fig:2} (bottom) represents the averaged density in a
sphere of $2$fm radius ($\rho _{_{2fm}}$) whereas the dashed line stands for
the density averaged in a sphere of $5$fm ($\rho _{_{5fm}}$).The sphere of
radius $5$fm is big enough to contain most of the volume of the nucleus and
small enough (see figure \ref{fig:1}) to not be polluted by the surface
shape. As expected, $\rho _{_{5fm}}$ presents a regular oscillation at the
breathing mode frequency. On contrary, it is clear that $\rho _{_{2fm}}$
contains more than the breathing mode, in particular, this quantity is more
sensitive to the presence of hole at the center of the nucleus as seen in
fig.\ref{fig:1}. Therefore, it appears that in addition to the expected
large amplitude volume mode other modes affecting both the interior and the
surface of the nucleus are present in a complete TDHF simulation.

\subsection{Collective vibrations}

\begin{figure}[tbph]
\begin{center}
\includegraphics[height=15cm,width=10cm]{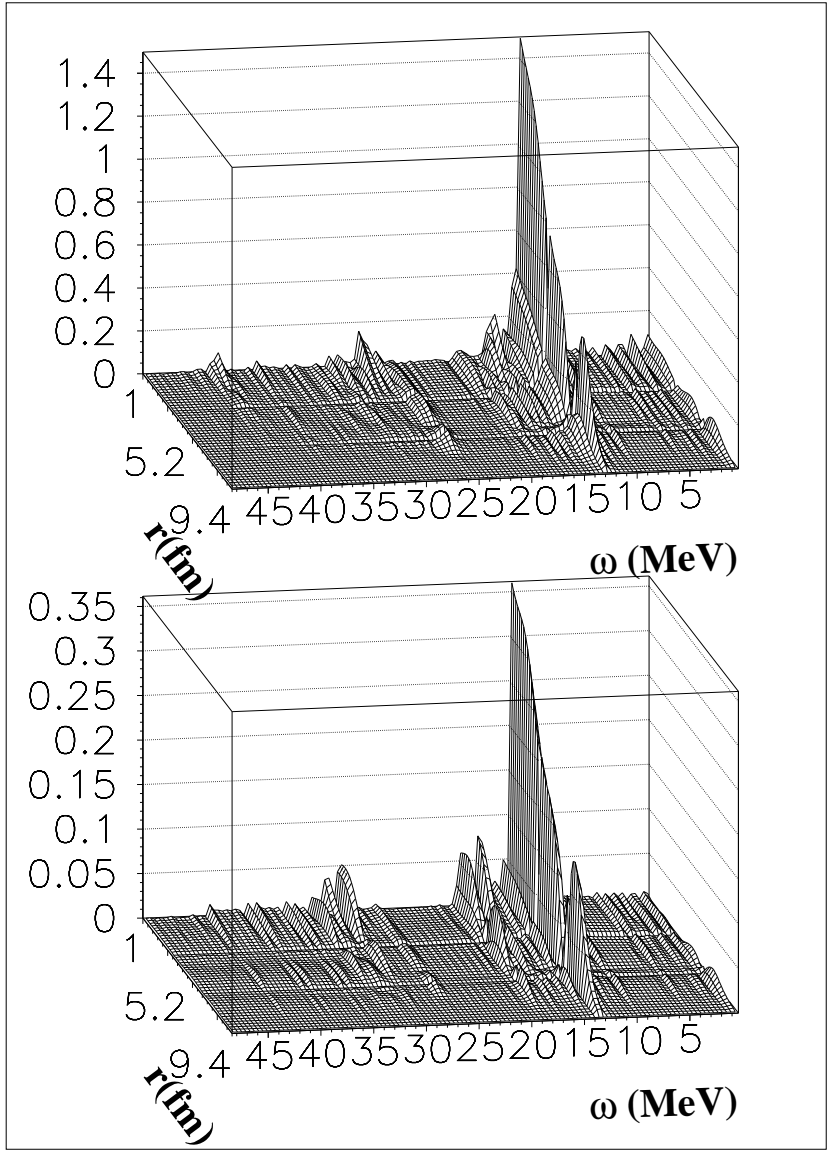}
\end{center}
\caption{Fourier transform $\left| \tilde{\rho}(r,\omega )\right| $ (3D
view). Two initial conditions are studied $\rho /\rho _{0}=0.8$ (top) and $%
\rho /\rho _{0}=1$ (bottom) are drawn. Fourier transforms are performed
over $1500 fm/c$.}
\label{fig:3}
\end{figure}

In order to quantitatively analyze the observed complex oscillation pattern
of the density, we have performed the Fourier transform of $\rho (r,t)$
leading to the spectral density $\tilde{\rho}(r,\omega )$. The Fourier
transform is performed up to $1500fm/c$ which is sufficient to have a
resolution of few hundred keV. A plot of $|\tilde{\rho}(r,\omega )|$ for $%
\rho /\rho _0=0.8$ is displayed in {fig. }\ref{fig:3}(Top). We see in
particular that non-local waves develop during the evolution. Of course we
recover the typical breathing mode at $12.5$ MeV but also contributions
around $17$MeV and $26$MeV. The radial dependence of $\tilde{\rho}(r,\omega )
$ for a given $\omega $ can be considered as the transition density of the
corresponding mode. Therefore, the appearance of hole at the center of the
nucleus seems to be connected with the surface waves. We have tested that
removing the coulomb field does not modify the presented conclusions (as
suggested in \cite{borderie}). Therefore, the creation of a hollow structure
is not a consequence of the coulomb interaction.

It should be noticed that starting from a value of $0.8$, $\rho
_{_{2fm}}/\rho _{_0}$ reaches much lower values during its evolution.
Moreover, the minimum minimorum (namely $\rho _{_{2fm}}/\rho _{_0}=0.4$) is
not reached during the first oscillation but at the third ones. The observed
oscillation pattern of $\rho _{_{2fm}}$ is characteristic of the beating of
several modes as clearly seen from the Fourier transform. It should be
noticed that reducing the amplitude of the oscillations we have not observed
a major modification of the Fourier spectrum. An example of such a case is
shown in figure 3 (Bottom) for initial compression $\rho /\rho _{_0}=1$. As
far as the main peaks are concerned, the frequency structure of the Fourier
transform does not change by more than a half MeV when the amplitude of the
mode increased. However, looking in details we can see that the relative
importance of the different modes is not the same. This is what we expect
since the amplitude of each mode is directly related to the initialization
procedure. This demonstrates that the coupling between modes and the
anharmonicities are small, and that the main effect is the beatings of
different modes. However, looking to the details of the radial structure of
the Fourier transforms we can notice that the transition densities are
slightly modified by the modification of the initial amplitude. This is
clearly evidenced in figure 4, where the transition densities $\tilde{\rho}%
(r,\omega )$ for the two main peaks $\omega =12$ MeV and $\omega =17$ MeV is
displayed for the two initializations shown in figure 3. In this figure we
can see that the Giant monopole resonance around $12$ MeV is more robust
than the smaller peak at $17$ MeV for which a contribution in the interior
of the nucleus is seen at large amplitude. This volume contribution explains
why a hollowed shape can appear as a function of time as the two modes are
beating one again the other with different nodal structures.

\section{Conclusion}

In the presented simulation, the beatings of different monopole modes
lead to very low densities in
the interior of the nucleus and to the formation of a hollow structure after
a long time (2 or 3 oscillation period). This tendency of excited nuclei to
expand in hollow structures have already been observed in several dynamical
calculations of the Vlasov type \cite{buble,borderie} however the new
feature shown by the presented quantum calculations is that this structure
gets more pronounced after several collective oscillations. This analysis
shows that this is due to the presence of several collective degrees of
freedom with a transition density affecting the interior of the nucleus.

This could change the discussion about the possible fragmentation of nuclei
and also about the involved time scales. Indeed, a possible interpretation
of multifragmentation is that heavy excited system could reach dynamically a
configuration in which the nucleus is unstable against a splitting in
pieces. This might be in particular the case for low density regions or for
hollow structures. In the framework of one collective variable picture,
after pre-equilibrium, the excited nucleus reaches the lowest density region
at the first turning-point of the collective expansion. This has given a
widely admitted time-scale of this phenomenon\cite{Cug1} and is now often
used to discuss break-up of a system\cite{Papp1}. If now we focus on $\rho
_{2fm}$, we see that a nucleus could oscillate several times before reaching
a configuration presenting density profiles with low density regions. If
this configuration is unstable we see that the nucleus could wait a long
time, of the order of $100fm/c$ for the second turning-point or even $%
200fm/c $ for the third, before initiating its fragmentation. It should be
noticed that indications in direction of this scenario have been recently
reported in hard photons study of Heavy-Ion collision\cite{Sch1}.

Note also that we have observed the same behavior for different nuclei with
higher temperatures. Moreover, we have applied the same analysis with a hard
EOS (SIII force), and the same observations have been made. Therefore, the
reported conclusion are rather generic. From the analysis we have related
this behavior to the existence of a volume contribution with several nodes
in the transition density. This, together with the Landau spreading, may
occur as generic features and explains that the presented dynamics was found
in many simulations.

In summary, we show that realistic TDHF calculations of expanding nuclei
cannot be reduced to the simple large amplitude breathing mode picture.
Generically, the observed dynamics presents the beating of several modes so
that it may happen that low density regions can be reached after several
global oscillations. This may have strong qualitative and quantitative
consequences on the dynamical expansion of nuclei: in particular, time-scale
could be strongly increased and exotic shapes can be reached.

{\bf Acknowledgments}

We thank K. Bennaceur and G. Martinez for helpful discussions and M. Colonna
for a careful reading of the manuscript.

\newpage

\newpage 

\newpage 

\newpage 
\begin{figure}[tbph]
\begin{center}
\includegraphics*[height=10cm,width=10cm]{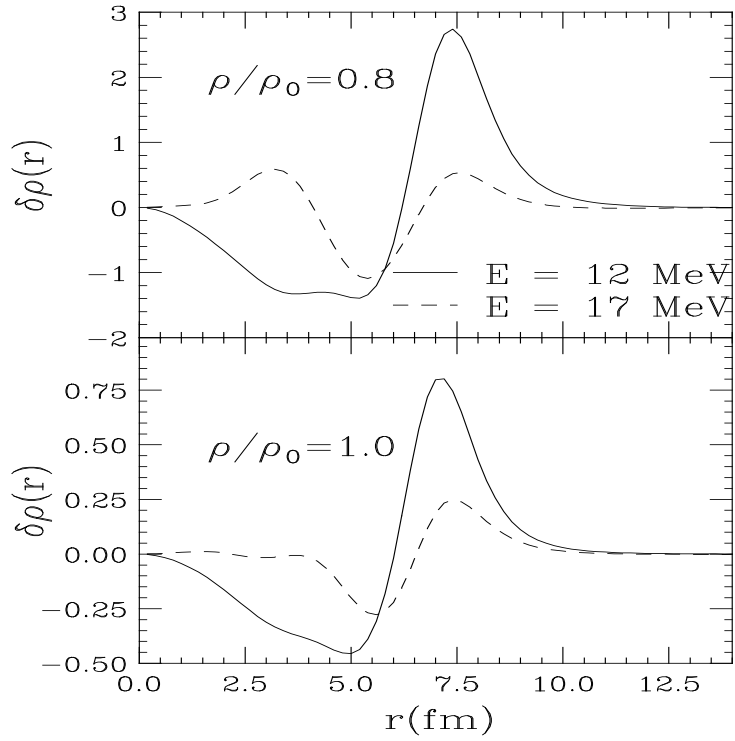}
\end{center}
\caption{Transition densities $\tilde{\rho}(r,\omega )$ for $\omega =12$ MeV
(solid line) and $\omega =17$ MeV (Dashed line) drawn for two values of the
initial compression (see fig 3).}
\label{fig:4}
\end{figure}

\newpage \medskip 

\end{document}